\magnification=1200
\input graphicx
\parindent=.7cm
\parskip=.25cm
\hoffset=20mm\voffset =20mm
\vsize=160mm
\hsize=125mm
%
\font\title=cmbx10 at 14pt
\font\eightrm=cmr8
\def\vec#1{{\bf #1}}
\def\fr#1{FR$_#1$}
\def\ph{\phantom{1}}
\def\today{\ifcase\month\or January\or February\or March\or %
                 April\or May\or June\or July\or August\or September\or %
                 October\or November\or Decembrer\fi \space\number\day,\space\number\year}

\line{\hfill LPSC-05-69}
\vskip -.1cm
\line{\hfill nucl-th/0508059}
\vglue 3cm
\centerline{\title Improving the Feshbach--Rubinow approximation}
\vskip .2cm
\vskip .5cm
\centerline{\bf Jean-Marc Richard}
\centerline{Laboratoire de Physique Atomique et Cosmologie}
\centerline{Universit\'e Joseph Fourier--CNRS-IN2P3}
\centerline{53, avenue des Martyrs, F-38026 Grenoble cedex, France}
\centerline{email: jean-marc.richard@lpsc.in2p3.fr}
\centerline{PACS:  21.45.+v, 03.65.Ge}
\centerline{\today}
\vskip 2cm
\centerline{\bf Abstract}
\vskip .1cm\noindent
The binding energy of three identical bosons is estimated by coupled differential equations which generalise the Feshbach--Rubinow approximation. This method turns out to be rather efficient, especially in the limit of vanishing binding. 
\vfill\break

\beginsection 1. Introduction\par
 There is no shortage of methods for solving the quantum-mechanical three-body problem: hyperspherical expansion, Faddeev equations, variational methods with sophisticated search for the free parameters, Monte-Carlo algorithms, etc. Each of them has merits and limitations. For instance, the hyperspherical expansion is a systematic scan of the Hilbert space, and can come in several variants such as ``potential harmonics'' or ``adiabatic''. However, it involves  a rather delicate evaluation of the matrix elements of the potential within the basis of harmonics.
 
 In the  50's, a cleaver approximation, hereafter referred to as \fr1, was proposed by Feshbach and Rubinow [1]. It consists of seeking the bound-state wave function of three identical bosons of mass $m$ as
 $$ \Psi(x, y,z)={u(r)\over r^{5/2}}~,\quad r={x+y+z\over 2}~,\eqno(1)$$
 where $x=|\vec{r}_2-\vec{r}_3|$, $y=|\vec{r}_3-\vec{r}_1|$ and $z=|\vec{r}_1-\vec{r}_2|$ measure the distances between particles. The S-wave ground state, indeed, depends on three scalar, translation-invariant, variables, which can be chosen as $x$, $y$ and $z$. The approximation consists of restricting to a function of their sum. Using astutely the stationnary properties of the Schr{\"o}dinger equation, Feshbach and Rubinow derived  the radial function obeyed by  the radial function $u(r)$ and energy $E$,
 $$ -u''(r)+{15\over 4 r^2} u(r)+ 2 m V_{\rm e}(r) u(r)= {14 m\over 15} u(r) E~,\eqno(2)$$
where the effective potential results from the projection of the interaction $V$,
$$ V_{\rm e}(r)=\int V(x,y,z)\, {\rm d}\tau~,\quad
\int{\rm d}\tau={8\over r^5}\int_0^r y \,{\rm d}y\int_{r-y}^r (2r-y-z)z\,{\rm d}z~.\eqno(3)$$

This method has been applied to a variety of problems, in particular in nuclear physics [2], and is still used occasionally. However, in the case of baryon spectroscopy [3], the energy  provided  by \fr1\   was found [4] not to be very accurate, as compared, e.g., to the hyperspherical expansion in the approximation of a single partial wave, the so-called hyperscalar approximation (HA).

The best convincing success of \fr1\  is perhaps in  the domain of loosely-bound systems as encountered  for instance in studying Borromean binding [5]. The \fr1\  approximation gives a good estimate of the coupling threshold%
\footnote{${}^a$}{\vtop{\baselineskip=8pt\hsize=11cm\eightrm\noindent$\scriptstyle g\ge g_3$ ensures a three-body bound state in the potential $\scriptstyle \sum_{i<j} v(|\vec{r}_i-\vec{r}_j|$,  where $\scriptstyle v$ is attractive or at least, contains attractive parts}}
 $g_3$, to be compared to the coupling threshold $g_2$ of two-body systems. 
 
In Fig.~1, a comparison is made of the three-body binding energy of the simple pairwise exponential potential $V(x,y,z)=g\sum_x v(x)$, with $v(x)=-\exp(-x)$, computed from \fr1\  and HA, as a function of the coupling $g$. The constituent mass is set to $m=1$. The \fr1\  method is clearly much better when the coupling threshold is approached, at $g\sim1.2$.

\centerline{\includegraphics[width=8.5cm]{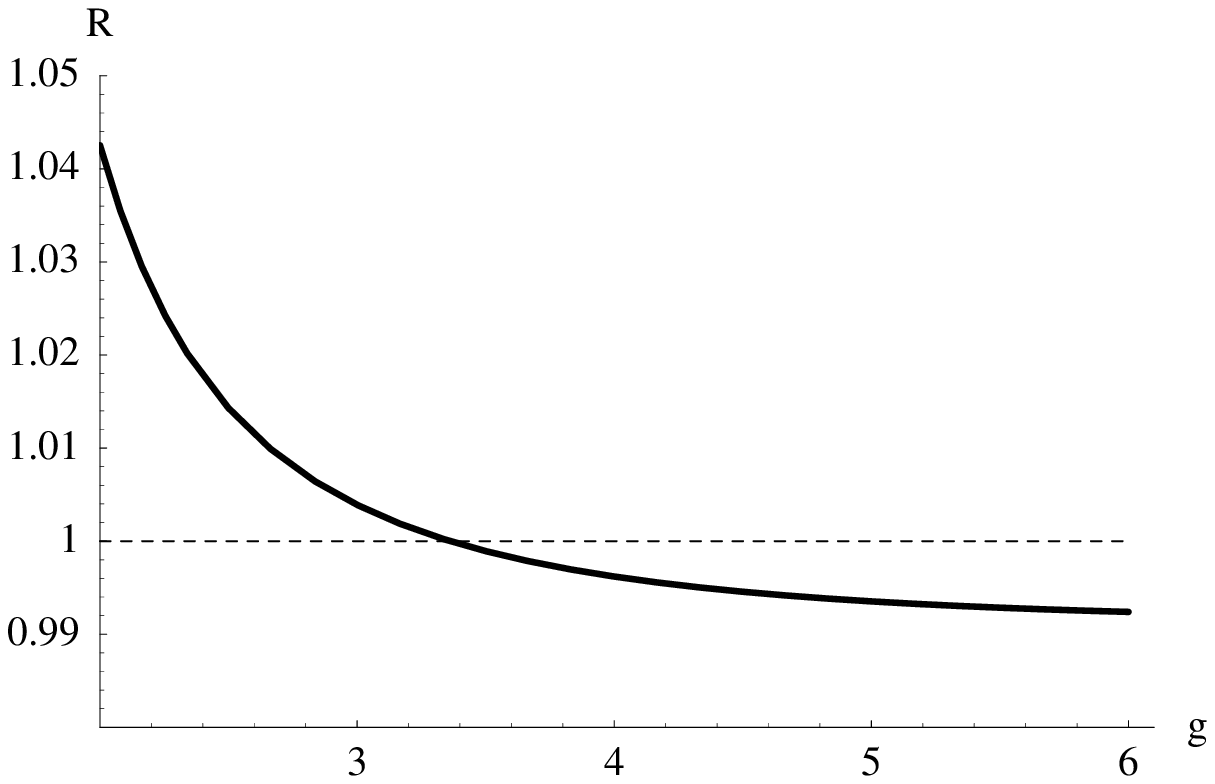}}
\centerline{\vtop{\hsize=8.7cm\noindent \sl Fig.~1: Ratio of Feshbach--Rubinow to hyperscalar approximations of  the three-body energy in a pairwise potential $-g \sum_{i<j} \exp(-r_{ij})$, as a function of the coupling~$g$.}}

\vskip .3cm

The \fr1\  method has been improved by Nogami et al. [2], Rosenthal et al.\ [6], and in particular adapted to deal with non-identical particles. 
In spite of its simplicity, the Feshbach--Rubinow method was never extensively used, probably because unlike the HA and other standard methods,  it was never presented as the starting point of a converging expansion. We shall suggest  below a possible remedy where the three-body wave function is written on a basis of orthogonal polynomials $P_i$ with weight functions $u_n(r)$ that generalise the single $u(r)$ of \fr1\  and obey coupled equations.

The ground-state energy obtained from these coupled equations will be compared to the value given by standard variational methods. For estimating the quality of the wave function, the short-range correlation coefficient $\langle\delta^{(3)}(\vec{r}_1-\vec{r}_2)\rangle$ will also be calculated. This quantity enters a number of decay or production rates, and is notoriously hard to compute accurately.
Is estimate is sometimes made easier by using identities involving the derivative of the potential and centrifugal terms [4]. We shall restrict ourselves in this paper to a direct reading of the wave function at $\vec{r}_1=\vec{r}_2$.
\beginsection 2. Formalism\par
The \fr1\  ansatz is generalised as \fr n\ 
 $$ \Psi(x,y,z)={u_1(r)\over r^{7/2}} P_1(x,y,z)+ {u_2(r)\over r^{9/2}}P_2(x,y,z)+\cdots~,\eqno(4)$$
 where $P_n$ is a real and homogeneous polynomial of degree $n$ which is symmetrical. The first one is simply $P_1=r$, and a truncation there corresponds to the original \fr1\  method. Among the  polynomials of degree $n=1$ in $x$, $y$ and $z$, only $P_1\propto x+y+z$, indeed,  is symmetric. For $n>1$,  the polynomials $r^{n-1}P_1$, \dots, $r P_{n-1}$ are already included in the first terms of the expansion (4), hence only  a few genuinely new polynomials have to be introduced.%
 \footnote{${}^b$}{\vtop{\baselineskip=8pt\hsize=11cm\eightrm\noindent For three variables ${\scriptstyle x}$, ${\scriptstyle y}$ and ${\scriptstyle z}$, the number ${\scriptstyle p}$ of symmetric polynomials of degree ${\scriptstyle n}$ does not exceed ${\scriptstyle n}$ till ${\scriptstyle n=5}$, and  ${\scriptstyle (p-n)/n}$ remains small for ${\scriptstyle n>5}$}}
     It is convenient to arrange the $P_n$ to be  orthogonal.
For $n=2$, the new polynomial is
 $$P_2(x,y,z)= \sqrt{3\over79}\left[ 72\, r^2-49 (x^2+y^2+z^2)\right]~,\eqno(5)$$
where the coefficient is chosen such that $\int{\rm d} \tau P_2^2=r^2\int{\rm d} \tau P_1^2$.
 
If  the expansion is truncated there, for testing purposes, the following set  of coupled equations can be derived by straightforward calculus:
 $$\eqalign{
 &\alpha_{11}u''_1+{\beta_{11}\over  r^2}u_1+\alpha_{12} u_2''+{\beta_{12}\over r^2} u_2+ \gamma_{12} {u'_2\over r} +V_{11} u_1+ V_{12} u_2={7\over 15} \epsilon u_1~,\cr
&\alpha_{22}u''_2+{\beta_{22}\over  r^2}u_2+\alpha_{21} u_1''+{\beta_{21}\over r^2} u_1+\gamma_{21} {u'_1\over r} + V_{21} u_1+ V_{22} u_2={7\over 15} \epsilon u_2~,}\eqno(6)$$
 with $\epsilon=2 m E$, $\alpha_{11}=-1$, $\alpha_{12}=\alpha_{21}=-2/(5\sqrt{237})$, $\alpha_{22}=-389/395$, 
 $\beta_{11}=15/4$, $\beta_{12}=-(93/10)\sqrt{3/79}$, $\beta_{21}=-95/(2\sqrt{237})$, 
 $\gamma_{11}=\gamma_{22}=0$, $\gamma_{12}=-\gamma_{21}=-98/(5\sqrt{237})$. The $2\times 2$ effective potential $V_{ij}$ results from the projections
 $$ V_{ij}=2m\int{\rm d}\tau\, P_i P_j V(x,y,z)~.\eqno(7)$$
 
 The coupled equations can be solved by several methods, for instance the discretisation procedure described in [4]. In short, $r\in[0,+\infty[$ is mapped into $x\in[0,1[$, and if $u_i(r)=v_i(x)$, each $v_i$ is expanded in Fourier series $v_i=\sum_n C_n \sin(n\pi x)$. The coefficients $C_n$ are translated into the values $v_{i}(x_n)$ of the radial fonctions at equally-spaced points $x_n=n/(N+1)$, where  $N$ is the number of points.  The system (6) is reduced to a $2N\times 2N$ matrix equation for the  quantities $v_{i}(x_n)$, and its eigenvalues give a good approximation to the energy levels of (6).
 
 The coefficient of short-range correlation,
 $$ \delta=\langle\Psi|\delta^{(3)}(\vec{x}|\Psi\rangle/\langle\Psi|\Psi\rangle~,\eqno(8)$$
 is given by
 $$
 \delta_1= {\displaystyle15\int_0^{+\infty}u^2 r^{-3}\,{\rm d r}\over \displaystyle7\pi\int_0^{+\infty}u^2 \,{\rm d r}}~,
 \quad
\delta_2={\displaystyle15\int_0^{+\infty}r^{-3}[u_1-26 u_2\sqrt{3/79}]^2\,{\rm d}r\over
\displaystyle7\pi\int_0^{+\infty}[u_1^2+u_2^2]\,{\rm d} r}~,
\eqno(9)$$
 for \fr1\ and \fr2, respectively.
\beginsection 3. Results\par
The method has been first tested on the harmonic oscillator, $v(x)=x^2$ for which the exact solution is 
$E=6\sqrt{3/2}\simeq 7.3485$. The  \fr1\  gives $E_1=(36/7)\sqrt{15/7}\simeq7.5284$, and the \fr2 leads to a remarkable improvement, with $E_2=7.3537$. Meanwhile, the ratio of short-range correlation $\delta$ to the exact value $\delta_{\rm ex}=(3/2)^{3/4}\pi^{-3/2}$ evolves from $\delta_1/\delta_{\rm ex}=1.57$ for \fr1\ to $\delta_2/\delta_{\rm ex}=0.968$, again a dramatic betterment.

The linear potential $\sum_{i<j}r_{ij}/2=r$ is diagonal in the basis of polynomials $P_i$, but is not exactly solvable because the kinetic energy is not diagonal. The results are $E_1\simeq3.906$ for \fr1\ and $E_2\simeq 3.8635$ for \fr2, to be compared with $E_{L=0}\simeq 3.8647$ and $E_{L\le4}\simeq 3.8633$ in the hyperspherical expansion with one and two partial waves, respectively [4]. The correlation coefficient is estimated to be $\delta_1=0.08488$ for \fr1\ and $\delta_2=0.05592$ for \fr2. For comparison, $\delta\simeq 0.05689$ is obtained from four coupled hyperspherical waves [4], and $\delta\simeq0.05702$ from 176 correlated Gaussians, a method to be presented shortly.

As for the short-range potentials, let us take again the example of the exponential potential $v(x)=-g \exp(-x)$ already considered in the introduction. Figure 2 displays the ratio of \fr2\ to \fr1\ ground-state energy as a function of the coupling $g$: the improvement saturates at about $2\%$ at large coupling and becomes more interesting in the domain of very weak binding. 

\vskip .2cm
\centerline{\includegraphics[width=5.5cm]{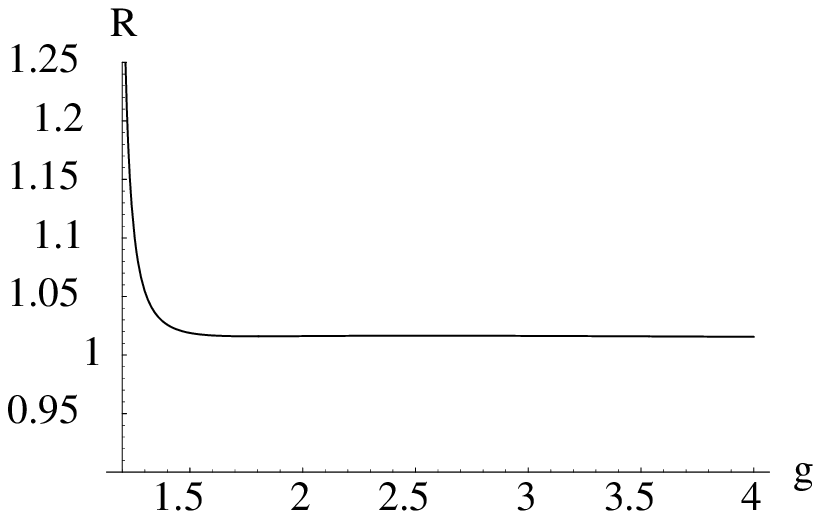}%
\hglue .7cm%
\vbox{\hsize=4.8cm\noindent \sl Fig.~2: Ratio of \fr2\ to \fr1 approximations of  the three-body energy in a pairwise potential $-g \sum_{i<j} \exp(-r_{ij})$, as a function of the coupling~$g$.}}
\vskip .2cm

In Fig.~3, our results are compared with these of one the fashionable methods on the market.
It consists of expanding the wave function on a basis of  correlated Gaussians
$$\Psi(x,y,z)=\sum_i C_i\left\{ \exp\left[-(a_i \vec{v}^2+2 b_i \vec{v}.\vec{w}+ c_i \vec{w}^2)/2\right]+\cdots\right\}~,
\eqno(10)$$
written in terms of the Jacobi variables 
$$\vec{v}=\vec{r}_2-\vec{r}_1~,\qquad \vec{w}={2\vec{r}_3-\vec{r}_1-\vec{r}_2 \over \sqrt3}~,
\eqno(11)$$
where the dots are meant for Gaussians deduced by permutation of the particles. Varga and Suzuki [7] have pushed  the method of stochastic search of the parameters to a high degree of  efficiency, and obtained very accurate estimates of binding energies in  a variety of domains. We shall use here the strategy of Hiyama et al.~[8]:  Gaussians with diagonal quadratic forms, i.e., $a_i \vec{v}^2+ c_i \vec{w}^2$, are first introduced, and if $a_i\neq c_i$, they are supplemented by the terms deduced by permutation, which includes an angular dependence through $\vec{v}.\vec{w}$ and carry the same weight factor $C_i$ to ensure Bose symmetry. The range parameters $a_i$ and $c_i$ are chosen to be any member of a geometric series $\{a, a r, a r^2, \ldots, a r^{N-1}\}$. This means that, however large is the number of terms, only two non-linear parameters, $a$ and $r$, are to be optimised. Note that $N=2$ range parameters correspond to $N'= 8$ Gaussians, $N=3$ to $N'=21$, $N=4$ to $N'=40$, i.e., a fairly large number of terms.

The plots in Fig.~3 indicate that
the \fr2\ value is better than this variational estimate with $N=2$, especially at vanishing binding, comparable to $N=3$, and is defeated shortly by $N=4$. This means that \fr2\ is roughly equivalent to an expansion involving 40 correlated Gaussians, at least for this particular potential.

\vskip .4cm
\centerline{\includegraphics[width=4.5cm,height=4cm]{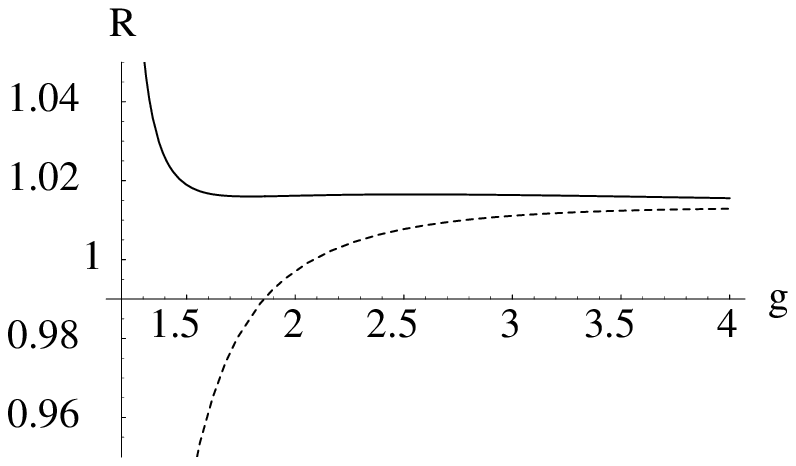}\includegraphics[width=4.5cm,height=4cm]{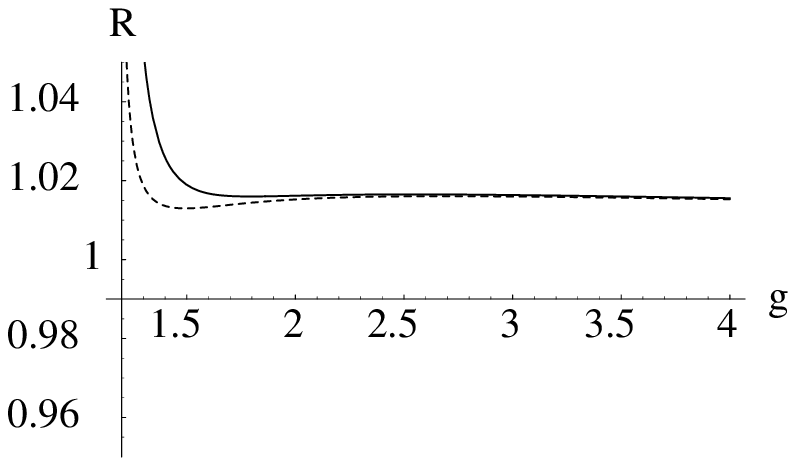}%
\includegraphics[width=4.5cm,height=4cm]{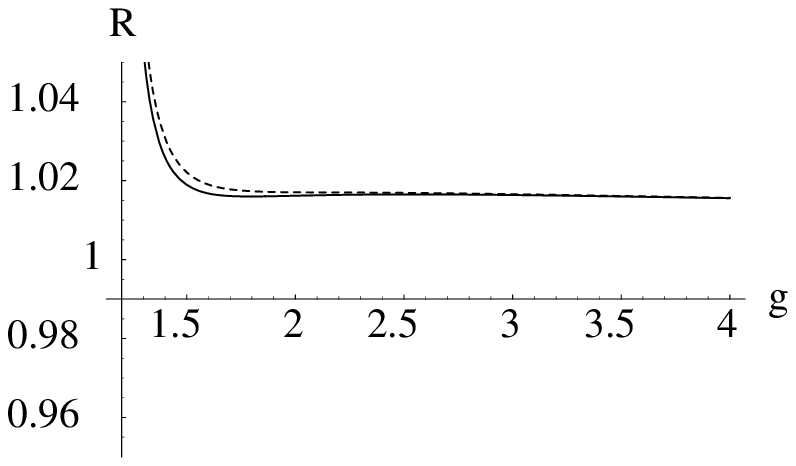}}
\centerline{\vtop{\hsize=10.5cm\noindent \sl Fig.~3: Comparison of the energy in the exponential potential, as a function of the coupling~$g$, for the \fr2\ approximation (solid line), and the Gaussian expansion (dashed line) with $N=2$ range parameters (left), $N=3$ (centre) and $N=4$ (right), corresponding to $N'=8,\ 20,\, 40$ terms. For clarity, the energies are shown by their ratio to the \fr1\ approximation.}}

For further checking, a basis of correlated exponential functions has also been used, namely
$$\psi(x,y,z)=\sum_i C_i\left[\exp(-a_i x-b_i y-c_i z)+\cdots\right]~,\eqno(12)$$
where the dots mean terms deduced by circular permutations of $\{a_i,b_i,c_i\}$. Wave functions of this type have been used in pioneering works on few-charge systems and in recent high-precision calculations of atomic systems [9]. The strategy of Ref.~[8] can also be applied, to avoid numerical instabilities in the minimisation process, by restricting all range parameters $a_i,b_i,\ldots$ to belong to a single geometric series. Note that with suitable constraints on these range parameters, this method can mimic the FR1 method ($a_i=b_i=c_i$ $\forall_i$) or its generalisation [6] by Rosenthal and Haracz  (RH), ($\{a_i,b_i,c_i\}\propto
\{a_1,b_1,c_1\}$ $\forall i>1$), $r^{5/2} u(r)$ being described as a sum of exponentials. This provides a check of the numerical solution of the radial FR1 equation or its RH analogue.

In Ref.~[6], indeed, Rosenthal et al.\ proposed to replace the symmetric ansatz  $F(r)=u(r)/r^{5/2}$, $2r=x+y+z$ of FR1 by 
$$\Psi(x,y,z)=F(\eta_1 x+\eta_2 y+\eta_3 z)+\cdots~,\eqno(13)$$
 where the dots correspond to permutation of $x$, $y$ and $z$ to restore the symmetry of the wave function. This method results in an integro-differential equation for $F$, whose solution is further optimised by varying  $\eta_2$ and $\eta_3$ empirically (a normalisation $\eta_1=1/2$ can be adopted).

In Table~1, a comparison is made of the values obtained for the ground-state energy and correlation coefficient in  the exponential potential at coupling $g=1.4$ and of the critical coupling $g_3$ at which three-body binding occurs. \fr1\ and \fr2\ are compared to RH, as well as to the Gaussian expansion $GN$ with up to $N$ different values of the range parameters, and it analogue $EN$ with exponentials.

At least with this potential, and for this domain of coupling close to the stability threshold, the generalisation of the Feshbach--Rubinow approximation with two coupled differential equations, \fr2, appears as more efficient than the single integro-differential equation RH. The ground-state energy, short-range correlation and coupling threshold approaches the exact value to about $1\%$ accuracy for the simple exponential potential.

 \vskip .2cm
 \centerline{\vtop{\hsize=11cm\noindent \sl Table 1. Ground-state energy $E_3$ and correlation coefficient $\delta$ for an exponential potential with coupling $g=1.4$, and critical coupling $g_3$ for three-body binding, with \fr1, \fr2 and  RH methods and variational calculations using correlated Gaussian ($G$) or exponential ($E$) basis functions with varying number of allowed range parameters.}}
 \vskip .2cm
 \centerline{%
 \vbox{\offinterlineskip
 \halign{\strut   
\hfil $ #\;$\hfil& 
\hfil $\; #\;$ \hfil&
\hfil $\; #\;$ \hfil&
\hfil $\; #\;$ \hfil&
\hfil $\; #\;$ \hfil&
\hfil $\; #\;$ \hfil& 
\hfil $\; #\;$ \hfil&
\hfil $\; #\;$ \hfil&
\hfil $\; #\;$ \hfil&
\hfil $\; #$ \hfil
\cr
 \noalign{\hrule}
 &G2&G4&G8&E2&E4&E5&\hbox{\fr1}&\hbox{RH}&\hbox{\fr2}\cr
  \noalign{\hrule}
 -E_3      &0.03055    &0.03575     &0.03586     &0.03466     &0.03583   &0.03586&0.03468   &
 0.03475    &0.03542\cr
\delta      &0.00840    &0.00972     &0.01000     &0.01251     &0.01005   &0.01002&0.01212   &
0.01224    &0.01091\cr
 g_3        &1.2194\ph &1.1613\ph &1.1563\ph  &1.2158\ph  &1.1587\ph&1.1560\ph&1.1751\ph&
 1.1649\ph&1.1644  \cr
 \noalign{\hrule}
 }}}

\beginsection 4. Outlook\par
In this paper, the Feshbach--Rubinow method is generalised by expanding the ground-state wave-function of three identical bosons on orthogonal polynomials of the interparticle distances. The coefficients of these polynomials  are functions of the sum of distances and obey  coupled equations which can be derived by calculus and solved numerically. The results obtained with two equations provide a dramatic improvement in the case of confining potentials where the single-equation version of this method was known to be inaccurate. For the short-range potentials, the results are also significantly better, and  become really excellent in the limit of loose binding,

Further investigations would be necessary to check the convergence when the number of equations is increased, and for excited states. The accuracy of the method should also be probed for potentials with hard core, which often occur in applications. The case of particles with different masses or interaction properties also deserves further work. The binding energy and wave function of  the positronium ion and other two-electron atoms, which were the subject of interesting developments of the Feshbach--Rubinow method,  will be studied in a forthcoming paper within the formalism of coupled equations.

\vskip .4cm\noindent
{\bf Acknowledgments:} I would like to thank Elian Masnada for stimulating discussions that led to this study, and 
Muhammad Asghar for very useful comments on the manuscript.
 
\beginsection References\par
\vskip -.2cm
\frenchspacing
\countdef\re=35
\re=0
\def\are{{\global\advance\re by 1}{\hbox to 4 mm{[\hfill\the\re]}}}
\parindent=.4cm\parskip=.05cm

\item{\are} H. Feshbach and S.I. Rubinow, Phys. Rev. {\bf 98}  (1955) 188.

\item{\are}   Leila Abou-Hadid and K. Higgins, Proc. Phys. Soc. {\bf 79} (1962)  34;
D. Maroun and M. McMillan,  Nucl. Phys. A  {\bf 159}  (1970) 661; R.K. Bhaduri and Y. Nogami, Phys. Rev. A {\bf 13}  (1976) 1986; S.G. Lie, Y. Nogami and M.A. Preston, Phys. Rev. A  {\bf 18} (1978) 787;
Zhongzhou Ren, J. Phys. G: Nucl. Part. Phys. {\bf 20}  (1994) 1185; Phys. Rev. C {\bf 49} (1994) 1281.

\item{\are} R.K.~Bhaduri, L.E.~Cohler and Y.~Nogami,
  Phys.\ Rev.\ Lett.\  {\bf 44} (1980) 1369.

\item{\are}  J.-M. Richard, Phys. Rep. {\bf  212} (1992) 1.
 
\item{\are}  C.M. Rosenthal and  R.D. Haracz,  J. Chem. Phys. 76, 1040 (1982); Phys. Rev. A {\bf 25} (1982)  1846.

\item{\are}   J. Goy, J.-M. Richard, and S. Fleck, Phys. Rev. A {\bf 52} (1995) 3511.

\item{\are}  Y. Suzuki, K. Varga, {\sl Stochastic variational approach to quantum mechanical few-body problems}, Springer-Verlag, (Berlin, Heidelberg) 1998. 

\item{\are} E. Hiyama, Y. Kino and M. Kamimura, Prog. Part. Nucl. Phys. {\bf 51} (2003) 223.

\item{\are}  See, e.g., H.A. Bethe and E.E. Sapleter, {\sl Quantum Mechanics of one- and two-electron atoms} (Plenum, New-York) 1977;
A.M. Frolov, Phys. Rev. A {\bf 69} (2004) 022505;
E.A.G. Armour et al., Phys. Rep. {\bf 413} (2005) 1; and refs.\  therein.

\end